\documentclass[twocolumn,trackchanges]{aastex701}

\usepackage{amsmath}
\usepackage{float}
\usepackage{threeparttable}
\usepackage{appendix}
\usepackage{booktabs}
\usepackage{multirow}

\newcommand{\swiftbat}[1]{\textit{Swift}-BAT}
\newcommand{\swiftxrt}[1]{\textit{Swift}-XRT}
\newcommand{\batxrt}[1]{\textit{Swift}-BAT/XRT}
\newcommand{\swiftUVOT}[1]{\textit{Swift}-UVOT}
\newcommand{\fermigbm}[1]{\textit{Fermi}-GBM}
\newcommand{\fermilat}[1]{\textit{Fermi}-LAT}

\accepted{April 20, 2026}

\begin{document}

\title{A Unified Framework for 10 TeV to EeV Diffuse Neutrino Sky and KM3-230213A}

\author[0000-0003-0035-7766]{Shiqi Yu}
\email[show]{shiqi.yu@utah.edu}
\affiliation{Department of Physics and Astronomy, University of Utah, Salt Lake City, Utah, USA}

\author[0000-0003-2478-333X]{B. Theodore Zhang}
\email[show]{zhangbing@ihep.ac.cn}
\affiliation{Key Laboratory of Particle Astrophysics and Experimental Physics Division and Computing Center, IHEP, CAS, China}
\affiliation{TIANFU Cosmic Ray Research Center, Chengdu, Sichuan, China}

\begin{abstract}
Establishing a unified framework that simultaneously accounts for the wideband diffuse neutrino flux and the physical origin of individual ultra-high-energy (UHE) neutrino detections, including KM3-230213A, remains a pressing challenge in multi-messenger astrophysics. In intrinsically low-luminosity gamma-ray bursts (LL~GRBs) driven by shock breakouts (SBOs), the evolving physical conditions naturally produce a multicomponent neutrino flux extending from 10 TeV to the EeV scale. By integrating prompt and afterglow phases within a unified framework grounded in multiwavelength observations of representative events, we show that LL GRB population accounts for this broadband neutrino emission through a characteristic two-hump spectrum. In this framework, the prompt emission from GRB~060218-like events accounts for $\gtrsim 10\%$ of the diffuse flux at 100~TeV, while GRB~100316D-like afterglow configuration predicts a distinct flux peak near $10^{-9}\rm~GeV~cm^{-2}~s^{-1}~sr^{-1}$ at 100~PeV.
This two-hump spectrum provides a high-energy component flux consistent with the 220 PeV KM3-230213A event, while the low-energy component contributes non-trivially to the observed diffuse neutrinos and supports the lack of individual low-energy counterparts. Furthermore, we utilize Fermi-LAT gamma-ray upper limits to place constraints on the source distance and luminosity of the event, assuming a GRB 100316D-like afterglow configuration. Ultimately, this framework identifies SBO-like LL~GRBs as a unifying origin for these phenomena, providing a physical link across the 10 TeV to EeV neutrino sky that is testable by next-generation observatories, including GRAND, IceCube-Gen2, and RNO-G.
\end{abstract}


\section{Introduction} \label{sec:intro}
The observed diffuse neutrino flux has been primarily characterized by IceCube at sub-PeV energies, revealing a structured spectral distribution that suggests a combination of multiple astrophysical source populations or distinct emission phases within individual transient classes \citep{Aartsen_2015,IceCube:2020acn,IceCube:2020wum,Abbasi_2022,naab2023measurementastrophysicaldiffuseneutrino,IceCube:2025tgp}. Two recent observational milestones have further redefined the neutrino landscape by providing critical new evidence regarding the diffuse sky. First, a new IceCube analysis has confirmed a broken power law structure in the diffuse neutrino energy spectrum with a significance level greater than 4~$\sigma$ \citep{IceCube:2020acn}. Second, the recent detection of an unprecedented 220 PeV neutrino event by KM3NeT \citep{KM3NeT:2025ccp} represents a new phenomenon that sits well above the energy range of previous IceCube observations \citep{Kurahashi:2022utm}. 

A promising candidate to explain both is the population of low-luminosity (LL) gamma-ray bursts (GRBs), which constitute a distinct population within the GRB family. While the LL~GRB classification can occasionally include off-axis observations of high-luminosity events, we focus here on the intrinsically low-luminosity population. Their high local event rate and mildly relativistic outflows, combined with the dense and compact environments of their progenitors, enable efficient cosmic-ray (CR) acceleration and enhanced photohadronic interactions \citep{Campana:2006qe, Soderberg_2006, Fan:2010br}. These properties distinguish them from high-luminosity (HL) GRBs, whose neutrino production efficiency is tightly constrained by IceCube analysis~\citep{Abbasi:2022whi}, and limited in contribution to the observed IceCube diffuse flux~\citep{Murase:2022vqf, Ai:2022kvd, Liu:2022mqe}.

Early studies~\citep{Murase_2006, Gupta:2006jm} proposed that LL~GRBs contribute non-negligibly to the observed diffuse neutrino flux. Subsequent works extensively explored particle acceleration and neutrino production across various scenarios. For the prompt phase, these include trans-relativistic shock breakouts~\citep{Kashiyama:2013qet}, choked jets with ultra-long GRBs~\citep{Murase:2013ffa}, red supergiants accompanied by Type-II supernovae \citep{He:2018lwb}, and short GRBs following binary neutron star mergers \citep{Gottlieb:2021pzr}. The contribution to diffuse neutrino flux was assessed in comprehensive studies in~\cite{Senno:2015tsn} across choked jet, shock breakout, and emergent jet models. Furthermore, a few studies discussed the role of LL~GRBs in explaining both ultra-high-energy cosmic rays (UHECRs) and diffuse neutrinos via heavy nuclei injection within internal shocks~\citep{Boncioli:2018lrv, Yoshida:2020div, Yoshida:2024fiu}. Regarding the afterglow phase, which is expected to provide a higher-energy contribution compared to the prompt phase, earlier works predominantly relied on HL GRB properties and hand-picked parameters, ignoring the full dynamical evolution of the system~\citep{Waxman:1999ai, Li:2002dw, Murase:2007yt}. A detailed modeling of the hadronic process in the LL~GRB afterglow phase from initial deceleration through the late-time non-relativistic regime, leveraging multi-wavelength observations, is currently lacking.

A recent systematic study of neutrino emission from seven LL~GRBs provided the first observational constraints on their physical parameters, including the CR loading factor $\xi_{\rm cr} \equiv L_{\rm cr}/L_{\gamma,\rm iso}$, the ratio between CR luminosity and isotropic-equivalent $\gamma$-ray luminosity~\citep{ref:PaperI}. It revealed an intrinsic diversity within the LL~GRB population and highlighted the necessity of isolating physically distinct subclasses. Among these, GRBs~060218 and 100316D are particularly representative as both are associated with supernova and exhibit shock-breakout (SBO) signatures~\citep{Kashiyama:2013qet} and similar energy partitioning among relativistic electrons, magnetic fields, and non-thermal protons.
This near-equipartition configuration naturally explains their prompt X-ray-dominated emission without invoking extreme microphysical parameters and suggests favorable conditions for efficient neutrino production. Their characteristic low bulk Lorentz factors and wide beaming angles provide a more robust basis for hadronic modeling compared to typical high-luminosity GRBs viewed off-axis~\cite{Zhang:2025abk}.

In this \textit{Letter}, we introduce a self-consistent, multi-zone framework. As the first attempt to explain the highest-energy neutrino detection within the context of the LL~GRB population, we constrain the physical parameters of the emission phases by leveraging afterglow multi-wavelength data. With the representative SBO-like LL~GRBs, we demonstrate a fundamental decoupling of the emission sites: while the prompt phase naturally accounts for the sub-PeV background observed by IceCube, the afterglow phase produces a distinct ``second hump" at $\sim 100$ PeV. This fills the gap between the sub-PeV structural diffuse flux and the UHE detection of KM3-230213A with a unified physical basis.

Throughout this work, we adopt a flat $\Lambda$CDM cosmology with $H_0 = 67.3\rm~km~s^{-1}~Mpc^{-1}$, $\Omega_M = 0.315$, and $\Omega_\Lambda = 0.685$~\citep{ParticleDataGroup:2024cfk}.

\section{Models and Methods}
We model the multi-messenger emission from LL~GRBs for both prompt and afterglow phases. Hadronic interactions and the resulting multi-wavelength photon and neutrino spectra are simulated using the numerical code \textsc{AMES}~\citep{2011ApJ...732...77M, Zhang:2023ewt, Zhang_2025}. 

\subsection{Models}\label{sec:model}
For the prompt phase, we assume a single bulk-shock scenario consistent with SBO or mildly relativistic jet dissipation. In the joint analysis presented here, we adopt the same physically motivated prior ranges as detailed in Section 2 of \cite{ref:PaperI} and summarized in Table~\ref{tab:prompt_bfp}. The following discussion is focused on the afterglow modeling.

The expansion of the mildly relativistic ejecta into the circumburst medium (CBM) drives an external forward shock (FS) and reverse shock (RS). While the RS can contribute to early-time emission, it is typically sub-dominant in the multi-wavelength observations of the events considered here. Therefore, in this work, we focus on the FS as the primary acceleration site for the conversion of bulk kinetic energy into non-thermal particles \citep{Sari:1995nm, Meszaros:1996sv, Sari:1997qe}. For GRB~060218 and 100316D, we adopt a wind-stratified CBM density profile ($n_{\rm CBM} = 3\times 10^{35} A_\star r^{-2}$)~\citep{Duran:2014naa}, where $A_\star$ is the ratio of the mass-loss rate to the wind velocity, normalized to $10^{-5} M_\odot \, {\rm yr}^{-1} / 10^3 \, \rm km \, s^{-1}$. We treat $A_{\star}$ as a free parameter with a flat prior range of $[0.01, 5]$, consistent with Wolf-Rayet progenitors \citep{Chevalier:1999mi}.

For mildly-relativistic jets, the deceleration radius is $R_{\rm dec} \simeq 2.7\times10^{15}\,\mathcal{E}_{k,51}\, \Gamma_{0,1}^{-2}\, A_\star^{-1}$~cm and the ejecta spreading radius is $R_s \approx \Delta_0 \Gamma_0^2 \simeq 3 \times 10^{15}\, T_{\rm ej,3}\, \Gamma_{0,1}^2$~cm~\citep{Zhang:2018ond}. 
The kinetic energy $\mathcal{E}_k$ is estimated from the isotropic luminosity $\mathcal{E}_k = \eta_\gamma^{-1} \times L_{\gamma,\rm iso}\times T_{90}$ with $\eta_\gamma = 0.1$ as the radiation efficiency. 
The condition $R_s \gtrsim R_{\rm dec}$ places these events in the thick-shell regime. We follow the standard evolution of the Lorentz factor and radius both during and after the shock crossing~\citep{Meszaros:1996sv,Sari:1997qe, 2013MNRAS.433.2107N}. 
When the shock system finishes traversing the ejecta at the shock-crossing radius, $R_\times \sim 1.6\times10^{15}\,\mathcal{E}_{k,51}^{1/2} T_{\rm ej,3}^{1/2} A_\star^{-1/2} \text{ cm}$, the emission from the shock system reaches its peak. 
We treat the initial bulk Lorentz factor $\Gamma_0$ as a free parameter with a flat prior range of $[2, 5]$. The lower bound ensures the eject expand at mildly relativistic velocities, while the upper bound encompasses the maximum speeds typically inferred for shock breakout in LL~GRBs 060218 and 100316D~\citep{2007ApJ...667..351W,Margutti_2013}. 

The magnetic field strength within the shocked medium at the shock crossing time is estimated as $B \approx 7.9\, \epsilon_{B,-4}^{1/2}$ G, assuming that a fraction $\epsilon_B$ of the post-shock internal energy density is partitioned into the magnetic field. We treat $\log_{10}\epsilon_B$ as a free parameter with a prior range of $[-5,\ -3]$, consistent with the values inferred from observations~\citep{Miceli:2022efx}. 

The maximum proton energy, limited by the dynamical timescale ($t_{\rm acc} \approx t_{\rm dyn}$), is $E_{p,\rm max}^{\rm dyn} \approx \eta^{-1}\,\Gamma_\times\, Z e B \left(R_\times/\Gamma_\times \right) \sim 3.8 \times 10^{17}\, \epsilon_{B,-4}^{1/2} \eta_1^{-1}\ {\rm eV}$, where $\eta$ is a numerical factor associated with the acceleration efficiency~\citep{Sironi:2015oza}. 

In the afterglow phase, we assume that a fraction $\epsilon_e$ ($\epsilon_p$) of the post-shock internal energy is carried by non-thermal electrons (protons) with a spectral index $s_e$ ($s_p$). We treat $\epsilon_e$ and $s_e$ as free parameters. The prior for $\epsilon_e$ is flat in log-space over the range $\log_{10} \epsilon_e \in [-2.0, -0.3]$, informed by systematic afterglow studies~\citep{Duran:2014naa,2011ApJ...726...75S,2014ApJ...785...29S}. The flat prior for $s_e$ is in the range of [2.05, 2.6], tailored to the observed properties of LL~GRBs 060218 and 100316D \citep{Campana:2006qe, 2011MNRAS.411.2792S}. Given the lack of direct observations to constrain proton acceleration, we adopt typical values of $\epsilon_p = 0.1$ and $s_p = 2$. 

The injection and cooling Lorentz factors are defined as:
\begin{equation}
\gamma_{e, \min} \approx \frac{\epsilon_e}{f_e} \frac{s_e-2}{s_e-1} \frac{m_p}{m_e} (\Gamma-1)
\label{eq:gmin}
\end{equation}
and $\gamma_{e, \rm cool} \approx 6\pi m_e c / [(1+Y_{\rm tot})\sigma_T \Gamma B^2 t]$. Here, $f_e$ denotes the fraction of accelerated non-thermal electrons. Although $f_e$ degenerates with other microphysics and normalization parameters, it represents the distinct physical process of particle injection efficiency and is therefore left free with a flat log-space prior of $\log_{10} f_e \in [-2.0, -0.3]$. This prior is motivated by previous studies~\citep{Duran:2014naa}, but incorporates an extended upper bound to remain conservative. The observed multi-wavelength data can meaningfully inform shock microphysics and ambient environment and mildly constrain parameters such as $f_e$. We summarize the full set of prior ranges and their corresponding posterior constraints in Table~\ref{tab:afterglow_bfp}.

To estimate the efficiency of photohadronic interactions and the resulting neutrino production, we adopt characteristic parameters for a typical LL~GRB. Assuming $\Gamma = 3$ and $A_\star = 1$, the electron luminosity is $L_e \approx \epsilon_e 4\pi R^2 c\Gamma^2 u_{\rm sh}^\prime \simeq 2.8 \times 10^{45} R_{16}$ erg s$^{-1}$, where $u_{\rm sh}^\prime$ is the comoving frame internal energy density. 
The resulting synchrotron luminosity is $L_{\rm syn} \sim \zeta_\gamma L_e \sim 10^{45}$ erg s$^{-1}$, with the radiated fraction of electron energy $\zeta_\gamma \sim (\gamma_{e, \rm cool}/\gamma_{e, \min})^{2-s_e} \sim 3$ for $\gamma_{e, \min} \sim 10^5$. The characteristic energy of the synchrotron emission, which peaks near $\gamma_{e, min}$, is $\varepsilon_m \approx (eB/2\pi m_e c) \gamma_{e, \rm min}^2 \Gamma \simeq 3.8\rm~eV$ for $B = 1\rm~G$ at $t = 10^3\rm~s$. The efficiency of neutrino production is then $f_{p\gamma}(E_p^\prime) \approx [2 \hat{\sigma}_{p\gamma} / (1+\alpha)] (R_{16}/\Gamma) n_\gamma (E_p^\prime/E_{p,0}^\prime)^{\alpha-1} \simeq 0.2 \eta_{p\gamma}[\alpha](E_p^\prime/E_{p,0}^\prime)^{\alpha-1}$~\citep{Zhang:2019htg}, where the effective cross section of photohadronic interactions is $\hat{\sigma}_{p\gamma} \simeq 0.7 \times 10^{-28}\rm~cm^2$, and $n_\gamma \approx L_{\rm syn} / 4\pi R^2 \Gamma^2 c \varepsilon_\gamma$ is the comoving photon number density. A numerical factor of $\eta_{p\gamma}[\alpha] = 2 / (1 + \alpha)$ accounts for the dependence on the photon spectral index $\alpha$. For the target photon field peaking at $\varepsilon_m \sim 3.8\rm~eV$, the comoving proton energy required for the $\Delta$-resonance is $E_p^\prime \approx 0.5 m_p c^2 \Gamma (\bar{\varepsilon}_\Delta / \varepsilon_m) \sim 3 \times 10^{17}\rm~eV$. In the observer frame, these protons reach $E_p = \Gamma E_p^\prime \sim 10^{18}\rm~eV$, producing neutrinos with observed energies $E_\nu \approx 0.05 E_p \sim 50\rm~PeV$. 

\subsection{Method}

Given that GRB~060218 and GRB~100316D exhibit similar prompt phase properties as SBO candidates, we perform a joint spectral fit to constrain the shared model parameters. We employ a Markov Chain Monte Carlo (MCMC) approach to sample the posterior probability distributions, utilizing the multi-wavelength observations (following the methods and observations detailed in~\cite{ref:PaperI}). The resulting parameter constraints are summarized in Table~\ref{tab:prompt_bfp}, while the confidence contours and parameter correlations are provided in Appendix~\ref{append:contours}.

Since the afterglow emission is sensitive to the unique external environment and initial ejecta properties of each burst, we perform separate fits for the FS evolution of GRB 060218 and GRB 100316D. We apply our MCMC framework to their respective multi-wavelength light curves, incorporating radio, X-ray, and gamma-ray observations. The resulting constraints are summarized in Table~\ref{tab:afterglow_data}, while the median predictions and associated 68\% credible intervals are shown in Fig.~\ref{fig:afterglow-lightcurve}. Further plots of posterior distributions are provided in Appendix~\ref{append:contours}, and the results are discussed next.

\begin{figure}[!tbh]
    \centering
    \includegraphics[width=\linewidth]{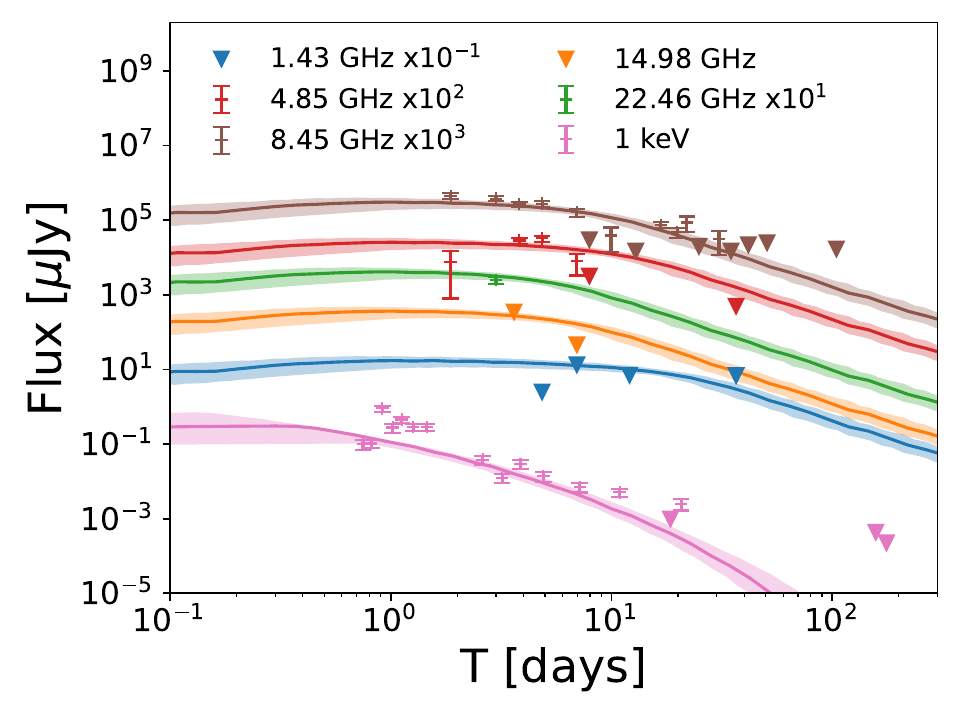}
        \includegraphics[width=\linewidth]{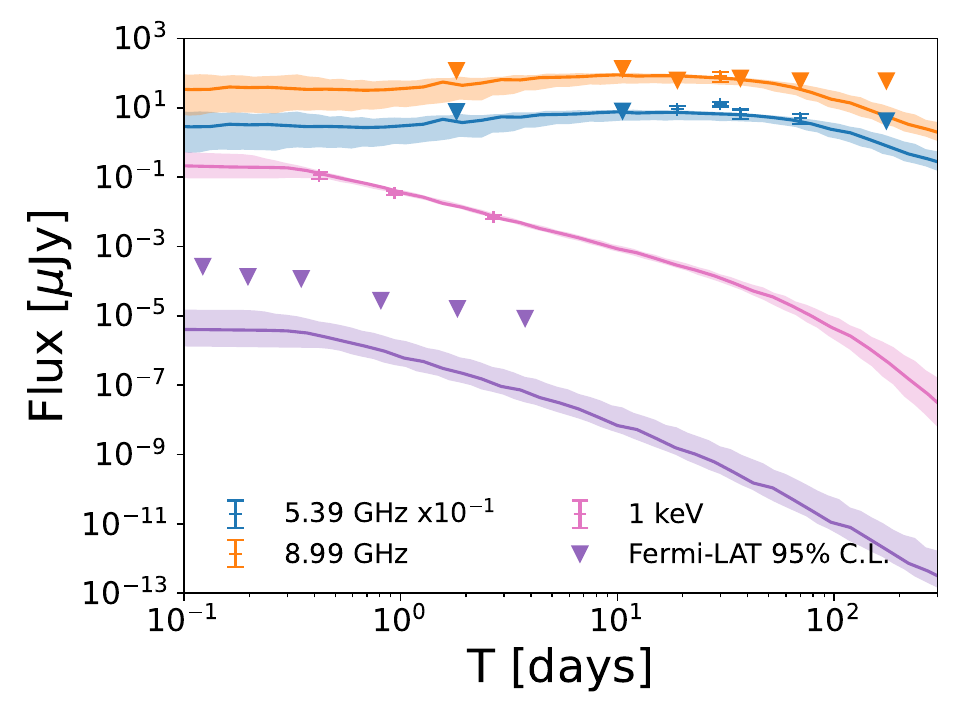}
    \caption{Multi-wavelength afterglow lightcurves for GRB 060218 (top) and GRB 100316D (bottom). Solid lines represent the posterior median of the FS model prediction, with 1$\sigma$ uncertainties (shaded), compared against observations (crosses) with 1$\sigma$ error bars or 95\% C.L. upper limits (triangles). The \fermilat{} upper limit is evaluated at 1 GeV.}
    \label{fig:afterglow-lightcurve}
\end{figure}

\section{Parameter Estimation}

\begin{table}[ht]
\centering
\caption{Inferred parameter values for the prompt phase model. Values represent posterior medians with 1D marginalized 1$\sigma$ highest posterior density (HPD) uncertainty ranges and corresponding ranges of flat priors.}
\begin{tabular*}{\columnwidth}{@{\extracolsep{\fill}} l c c c }
\hline
Parameter & Median & 1$\sigma$ uncertainty & Prior range \\
\hline
$\log \xi_B$ & 2.0 & [1.62, 3.00] & [-3.0, 3.0] \\
$\log \xi_e$ & 0.8 & [0.51, 1.15] & [-0.5, 2.0] \\
$\log (R/[\rm{cm}])$  & 13.0 & [11.50, 13.84] & [11.5, 16] \\
$\Gamma_0$ & 7.9 & [2.00, 11.14] & [2.0, 20.0] \\
$s_e$ & 2.4 & [2.02, 2.59] & [2.0, 3.0] \\
$\epsilon_e$ & 0.1 & [0.01, 0.20] & [0.01, 0.3] \\
$\log \xi_p$ & -0.4 & [-2.00, 0.25] & [-2.0, 2.0] \\
\hline
\end{tabular*}
\label{tab:prompt_bfp}
\end{table}

\begin{table}[ht]
\centering
\caption{Afterglow model parameters for GRB~060218 and GRB~100316D. Values represent posterior medians with 1D marginalized 1$\sigma$ HPD credible intervals; flat prior ranges are provided for reference.}
\begin{tabular*}{\columnwidth}{@{\extracolsep{\fill}} l c c c }
\hline
Parameter & 060218 & 100316D & Prior range \\
\hline
$\log(\epsilon_e)$  & $-1.2^{+0.25}_{-0.33}$ & $-0.7^{+0.37}_{-0.14}$ & $[-2.0, -0.3]$ \\
$\log(\epsilon_B)$  & $-3.4^{+0.50}_{-0.45}$ & $-3.5^{+0.31}_{-0.35}$ & $[-5.0, -0.3]$ \\
$s_e$ & $2.1^{+0.01}_{-0.02}$ & $2.3^{+0.06}_{-0.06}$ & $[2.05, 2.6]$ \\
$\Gamma_0$ & $3.2^{+0.46}_{-0.70}$ & $2.3^{+0.16}_{-0.24}$ & $[2.0, 5.0]$ \\
$\log(f_e)$ & $-1.8^{+0.07}_{-0.18}$ & $-1.7^{+0.18}_{-0.15}$ & $[-2, -0.3]$ \\
$A_*$ & $0.5^{+0.33}_{-0.45}$ & $4.3^{+0.67}_{-0.34}$ & $[0.01, 5]$ \\
\hline
\end{tabular*}
\label{tab:afterglow_bfp}
\end{table}

The neutrinos produced during the prompt and afterglow phases are in different energy regimes. In the external FS region, the larger emission radius leads to lower magnetic field strengths and target photon densities compared to the prompt phase. This environment allows particles to reach the UHE regime more efficiently than in internal shock models \citep{PhysRevLett.78.2292}. 

For the prompt model, the joint fit reflects the shared physical properties of the SBO-like LL~GRBs. The electron spectral index, $s_e \sim 2.4$, is consistent with the diffusive shock acceleration (DSA) scenario~\citep{Sironi:2015oza}, while the energy partition fractions for electrons and magnetic fields ($\log \xi_e \approx 0.8$, $\log \xi_B \approx 2.0$) suggest near-equipartition and efficient particle acceleration, in agreement with the previous conclusion in~\cite{ref:PaperI}. The CR loading parameter, $\log \xi_p \approx -0.4$, is associated with the normalization of the neutrino production during the prompt emission phase. Given the current absence of direct gamma-ray and neutrino detections, the parameter constraints derived from Fermi-LAT upper limits represent an optimistic scenario for the prompt phase of LL GRBs. This interpretation ensures that the predicted neutrino flux is understood as a potential upper bound rather than an expected detection under current observational constraints.

The initial bulk Lorentz factor in our prompt analysis, $\Gamma_0$, is weakly constrained with a median of $7.9$ and a $1\sigma$ range of $[2.00, 11.14]$, as expected under the bulk-shock scenario. However, multi-wavelength light curves in the afterglow phase provide a better constraint to break the degeneracy between the emission radius and the expansion velocity. We find $\Gamma_0$ values of $2.3^{+0.16}_{-0.24}$ for GRB~100316D and $3.2^{+0.46}_{-0.70}$ for GRB~060218, consistent with the mildly-relativistic expansion speeds expected for SBO driven LL~GRBs. Additionally, our results yield comparible energy partition fractions for both bursts ($\log\epsilon_e\approx-1$, $\log\epsilon_B\approx-3.5$), which align with standard FS models for LL~GRBs~\citep{Miceli:2022efx}. The spectral index of accelerated electrons agrees with the DSA scenario ($2.1^{+0.01}_{-0.02}$ for 060218 and $2.3^{+0.06}_{-0.06}$ for 100316D).
Furthermore, the inferred CBM densities, $A_{\star} \approx 0.5^{+0.33}_{-0.45}$ for GRB 060218 and $A_{\star} \approx 4.3^{+0.67}_{-0.34}$ for GRB 100316D, are consistent with the ranges expected for the Wolf-Rayet star progenitors typically associated with LL~GRBs \citep{Campana:2006qe, Margutti_2013}.

Our results suggest that while the internal conditions of the sources determine the prompt emission, the afterglow evolution reflects the diversity in CBM densities and ejecta energetics. Next, we discuss how the two phases of SBO-like LL~GRBs contribute to the diffuse neutrino sky.

\section{Contribution to diffuse neutrinos and KM3-230213A event}\label{sec:diffuse}

The cumulative diffuse neutrino flux from the LL~GRB population is:
\begin{align}
    E_{\nu}^2 \Phi_{\nu} &= \frac{c}{4\pi H_0} \int_{z_{\min}}^{z_{\max}} dz \int_{L_{\min}}^{L_{\max}} dL_{\gamma} \frac{1}{(1 + z)^2 }\nonumber \\ &\times \frac{d\mathcal{\rho}_{\rm LL}(z)/dL_{\gamma}}{\sqrt{\Omega_M (1 + z)^3 + \Omega_\Lambda}} \left({E'}_{\nu}^2  \frac{dN_{\nu}}{d{E'}_{\nu}} \right),
\end{align}
where $dN_{\nu}/d{E'}_{\nu}$ represents the neutrino spectrum evaluated at the posterior medians of the model parameters and ${E'}_{\nu} = (1 + z) E_\nu$. For the afterglow phase, the neutrino fluence is integrated from $T = 1000\,\text{s}$ to $T = 3 \times 10^7\,\text{s}$ to capture the full evolution of the FS. We adopt redshift limits of $z_{\rm min}=0.001$ and $z_{\rm max}=5$, with a luminosity range of $L_{\rm min}=10^{46}$ to $L_{\rm max}=10^{50}\,\text{erg}\,\text{s}^{-1}$.

We adopt the luminosity function (LF) for SBO-like LL~GRBs as a single power law with a spectral index of 0.84 and a local event rate of $\rho_0 \sim 164\,\text{Gpc}^{-3}\,\text{yr}^{-1}$ for luminosities above $5 \times 10^{46}\,\text{erg}\,\text{s}^{-1}$~\citep{Sun:2015bda}. This choice is intended to represent the intrinsically faint population rather than typical GRBs viewed off-axis. Alternative LFs~\citep{Liang:2006ci,Sun:2022htd} have a negligible impact on the diffuse flux. Recent detections by the \textit{Einstein Probe}~\citep{2025arXiv250316243H} suggest a sample of SBO candidates that will further refine these population statistics and the resulting neutrino flux predictions.

\begin{figure}
    \centering
\includegraphics[width=\linewidth]{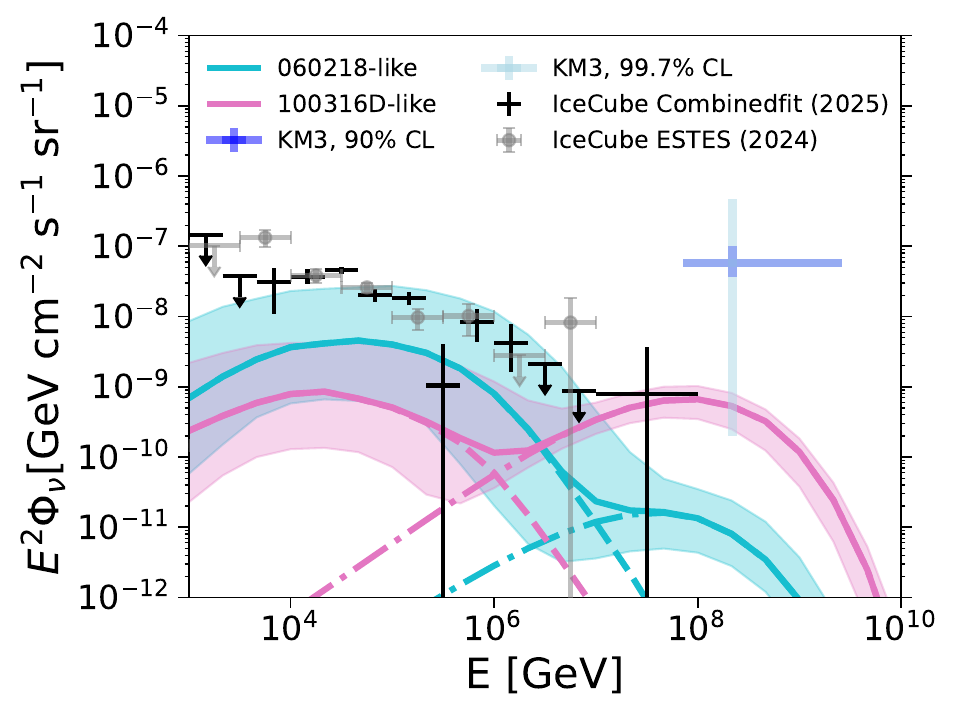}
\caption{Predicted diffuse neutrino flux (solid lines) with 1$\sigma$ uncertainties (shaded) from the SBO-like LL~GRB population compared to observed astrophysical neutrino fluxes (crosses)~\citep{IceCube:2024fxo,IceCube:2025tgp,KM3NeT:2025ccp}. Scenarios assume a population composed of either GRB~060218-like (blue) or GRB~100316D-like (pink) sources. The prompt (dashed) and afterglow (dotted-dashed) sub-components are shown separately, highlighting the distinct contributions to the two-hump spectral distribution.}
    \label{fig:GRB-diffuse}
\end{figure}

The resulting diffuse neutrino flux spectrum (see Fig.~\ref{fig:GRB-diffuse}) exhibits a characteristic two-hump structure, originating from the distinct physical regimes of the prompt and afterglow phases. In an optimistic scenario, the prompt emission can account for $\gtrsim$10\% of the diffuse flux at 100~TeV for a GRB 060218-like configuration, whereas the UHE component is strongly dependent to the circumburst environment. For instance, the dense stellar wind inferred for GRB 100316D enhances hadronic interaction efficiency, yielding a significantly higher afterglow-to-prompt flux ratio compared to the GRB 060218-like configuration. This second hump attains a flux level sufficient to explain the highest energy neutrino detected to date, KM3-230213A.

The detection of this 220 PeV event by KM3NeT poses a challenge for standard astrophysical source models, which struggle to explain the production of such a high-energy neutrino without companion observations \citep{KM3NeT:2025npi,KM3NeT:2025ccp,KM3NeT:2025lly,vx8j-pxy6}. Our framework naturally accommodates this event, as neutrinos can be produced during the afterglow phase while the prompt signal remains below detection thresholds, consistent with sources that are either intrinsically faint or cosmologically distant.

Since KM3-230213A lacks an electromagnetic counterpart, we do not expect to constrain the physical parameter space. Instead, we assume that the neutrino originates from the afterglow emission and adopt the representative physical parameters of GRB 100316D (from Table \ref{tab:afterglow_bfp}). Its afterglow neutrino fluence is significantly higher than that of GRB 060218, making it a more efficient producer of UHE neutrinos. We assume a prompt duration of 1 ks and a jet onset at $T_0 - 3$ ks. By fixing these physical properties and having only the distance and luminosity free, we use the observed \textit{Fermi}-LAT upper limits to constrain the allowed region where the SBO-like LL GRB assumption satisfies both the UHE neutrino detection and the multi-wavelength non-detection. The resulting region and the representing predictions of $\gamma$-ray flux along the boundary are illustrated in Fig. \ref{fig:km3_fit}, with the corresponding parameter space detailed in Appendix \ref{append:contours}.

\begin{figure}
\centering
\includegraphics[width=\linewidth]{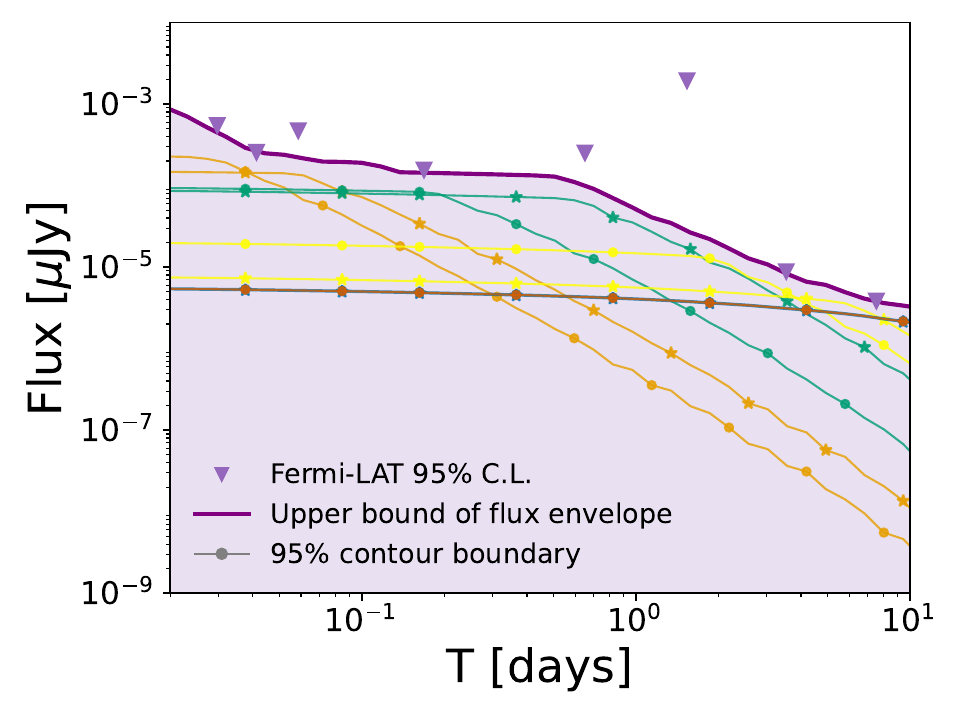}
\caption{Predicted gamma ray flux representing the post-burn-in posterior parameter space for a hypothetical SBO-like LL~GRB. The solid purple curve highlights the maximum predicted flux compared against the observed 95\% confidence level upper limits from \fermilat{}. The colored curves represent the parameter points along the allowed parameter space boundary, with colors and markers matching those in Fig.~\ref{fig:prompt_contour}.}
\label{fig:km3_fit}
\end{figure}

Thus, the KM3-230213A event could represent a rare, stochastic detection from a faint population of SBO-like LL~GRB. Despite the larger effective area of IceCube, the low flux produced during the afterglow phase ensures that a single detection in KM3NeT remains statistically compatible with null results elsewhere. While the contribution of off-axis events to the diffuse neutrino flux remains beyond the scope of this work, a comprehensive follow-up study would serve as a valuable future extension.

\section{Summary and Discussions}
This work presents the first unified framework to explain the 220 PeV neutrino event as a rare detection from the intrinsically faint SBO-like LL~GRB population. Our results demonstrate that the afterglow phase can produce such an orphan UHE neutrino while the prompt signal remains below detection thresholds, providing a physical explanation for the absence of coincident sub-PeV or electromagnetic counterparts.

Using multi-wavelength observations of the SBO-like LL GRBs 060218 and 100316D, we constrain the physical parameter spaces for both phases and demonstrate that this subclass can naturally produce a two-hump neutrino spectrum spanning from $10$~TeV to EeV, deviating from a single power law. Moreover, the non-power-law nature of the spectrum allows a source in a low-density environment to produce a detectable prompt signal without a corresponding UHE afterglow component. We also show that the two sources represent distinct regimes. The GRB 060218 configuration provides a significant flux of $\sim$100~TeV neutrinos during the prompt phase but lacks a substantial UHE component. Conversely, the GRB 100316D configuration contributes effectively to the UHE neutrino sky while remaining inefficient below 1~PeV. To contribute significantly to the entire diffuse neutrino flux, the population must exhibit a wide range of physical parameters, particularly the CBM properties that govern the UHE regime. Notably, our model requires a CR energy budget of $\mathcal{E}_{\rm CR} \sim 10^{50}\rm~erg$ per burst, below the $\sim 4 \times 10^{51}\rm~erg$ observational constraints of UHECR flux given the local LL~GRB rate~\citep{Zhang:2017moz, PierreAuger:2024flk}. While our analysis focuses on SBO-like LL~GRBs, these results establish a broader physical principle: any transient hosting successive shock regimes can naturally produce a multi-hump neutrino spectrum, such as stripped-envelope supernovae~\citep{Sawada:2024rfz}.

Using the publicly available effective areas for GRAND~\citep{GRAND:2018iaj}, IceCube-Gen2~\citep{IceCube-Gen2:2021rkf, Glaser:2019cws}, and RNO-G~\cite{RNO-G:2020rmc}, we estimate the integrated detection rates of afterglow phase over a 10-year period (see Table~\ref{tab:neutrino_sensitivity}).

\begin{table}[htb]
\centering
\caption{Projected number of events for future detectors: GRAND, IceCube-Gen2, and RNO-G over 10-year livetime for GRB~060218 and GRB~100316 configurations. 
}
\label{tab:neutrino_sensitivity}
\begin{tabular}{llccc}
\toprule
\toprule
Config. & Detector & Medium & Low & High \\
\midrule
\multirow{3}{*}{060218} & GRAND & 0.283 & 0.096 & 0.849 \\
                        & IceCube-Gen2 & 0.093 & 0.030 & 0.342 \\
                        & RNO-G & 0.002 & 0.001 & 0.007 \\
\midrule
\multirow{3}{*}{100316D} & GRAND & 19.467 & 8.674 & 29.711 \\
                        & IceCube-Gen2 & 5.547 & 2.564 & 8.532 \\
                        & RNO-G & 0.105 & 0.049 & 0.162 \\
\bottomrule
\end{tabular}
\end{table}

We note that the peak neutrino fluence from a single external FS is approximately $10^{-6}\rm~GeV~cm^{-2}$, which is roughly five orders of magnitude below the point-source sensitivity thresholds of upcoming radio-based UHE experiments~\citep{Kotera:2025jca}. Consequently, while individual event associations remain challenging, the next generation of large-scale neutrino telescopes will be uniquely positioned to test this model by searching for the predicted multi-component spectral signature in the UHE regime. The future of this paradigm depends on expanding the sample of soft X-ray transients through facilities like the Einstein Probe~\citep{Yuan:2025cbh}. Rapid multi-wavelength follow-up, particularly in the radio band, can be used to estimate the wind density of the CBM. Furthermore, future neutrino telescopes will reach the sensitivity required to reveal the second UHE hump, clarifying the role of LL~GRBs as promising candidates for the origin of high-energy neutrinos and cosmic rays.

\begin{acknowledgments}

We thank the anonymous referee for the invaluable comments. We thank Vedant Basu, Qi Feng, Dan Hooper, Tanmoy Laskar, Charles Jui, Carsten Rott, Wayne Springer, Xiangyu Wang, Chris Weaver, Shigeru Yoshida, Xuehui Yu, and Lihua Zhou for discussions and suggestions. 
B.T.Z. is supported in China by the National Key R\&D Program of China under the grant 2024YFA1611402.
We acknowledge the Center for High Performance Computing (CHPC) at the University of Utah for providing computational resources.
Correspondence and requests for materials should be addressed to S.Y and B.T.Z.
\end{acknowledgments}

\onecolumngrid
\appendix

\section{Multi-wavelength Observations}\label{append:data}

Observations for the prompt phase are identical to those used in \cite{ref:PaperI}. For the afterglow phase, we use multi-wavelength data from radio, X-ray, and $\gamma$-ray observations, summarized in Table~\ref{tab:afterglow_data}.

For GRB~060218, only the measured fluxes at 4.86, 8.46, and 22.5~GHz from VLA \citep{Soderberg_2006} are included in the fit, while all upper-limit points, including 1.43~GHz, are shown in the plots for comparison and completeness (see Fig.~\ref{fig:afterglow-lightcurve}). For GRB~100316D, ATCA radio data at 5.4 and 9~GHz \citep{Margutti_2013} are used in the analysis. X-ray light curves are obtained from the \textit{Swift}-XRT repository~\citep{Evans_2009}\footnote{\url{https://www.swift.ac.uk/xrt_spectra/}} when available.

For GRB~100316D and KM3-230213A, archival \fermilat{} $\gamma$-ray data (1–300~GeV) are analyzed using the \texttt{fermipy} (v1.4.0) framework. We select \texttt{evtype==1} (FRONT) events within a 20$^\circ$ region of interest (ROI) centered on the target location. To ensure a robust background model, we first perform a broadband likelihood fit over a 20-day baseline following the trigger time. During this optimization, the normalizations of the Galactic (\texttt{gll\_iem\_v07}) and isotropic (\texttt{iso\_P8R3\_SOURCE\_V3\_v1}) diffuse components, as well as all point sources within 5$^\circ$ (including the target), are left free. Subsequently, the afterglow light curve is extracted with all background parameters fixed and the target photon index fixed at $\Gamma = 2$. For GRB~100316D, the analysis window spans $T_{90}+3\,$ks to 5~days post-trigger to exclude prompt emission. For KM3-230213A, the light curve analysis starts at $T_0 - 1$ ks to isolate the afterglow-dominated regime, extending to $T_0 + 10$ days for broader temporal coverage. Given no significant signals are detected in any interval ($TS < 9$), we extract 95\% confidence level upper limits for both events. 

\begin{table}[ht]
\centering
\caption{Multi-wavelength observations used for afterglow fitting of GRBs~060218 and 100316D.}
\begin{tabular}{lcccc}
\hline
Band & GRB~060218 & GRB~100316D & Telescope & Reference\\
\hline
\multirow{6}{*}{Radio} 
 & 1.43 & -- & VLA & \multirow{4}{*}{\cite{Soderberg_2006}}\\
 & 4.86 & -- & VLA & \\
 & 8.46 & -- & VLA & \\
 & 22.5 & -- & VLA & \\
 & -- & 5.4 & ATCA & \multirow{2}{*}{\cite{Margutti_2013}}\\
 & -- &  9.0 & ATCA & \\
\hline
X-ray  & 0.3--10~keV  & 0.3--10~keV & \textit{Swift}-XRT & \multirow{2}{*}{\cite{Evans_2009}} \\
 & 00191157053 & 00416135019 & Observation ID & \\
\hline
$\gamma$-ray & -- & 1--300~GeV & \fermilat{} & \cite{Abdollahi_2022}\\
\hline
\end{tabular}
\label{tab:afterglow_data}
\end{table}

\section{Contour plots}\label{append:contours}

\begin{figure*}[hbt] \centering \includegraphics[width=0.55\linewidth]{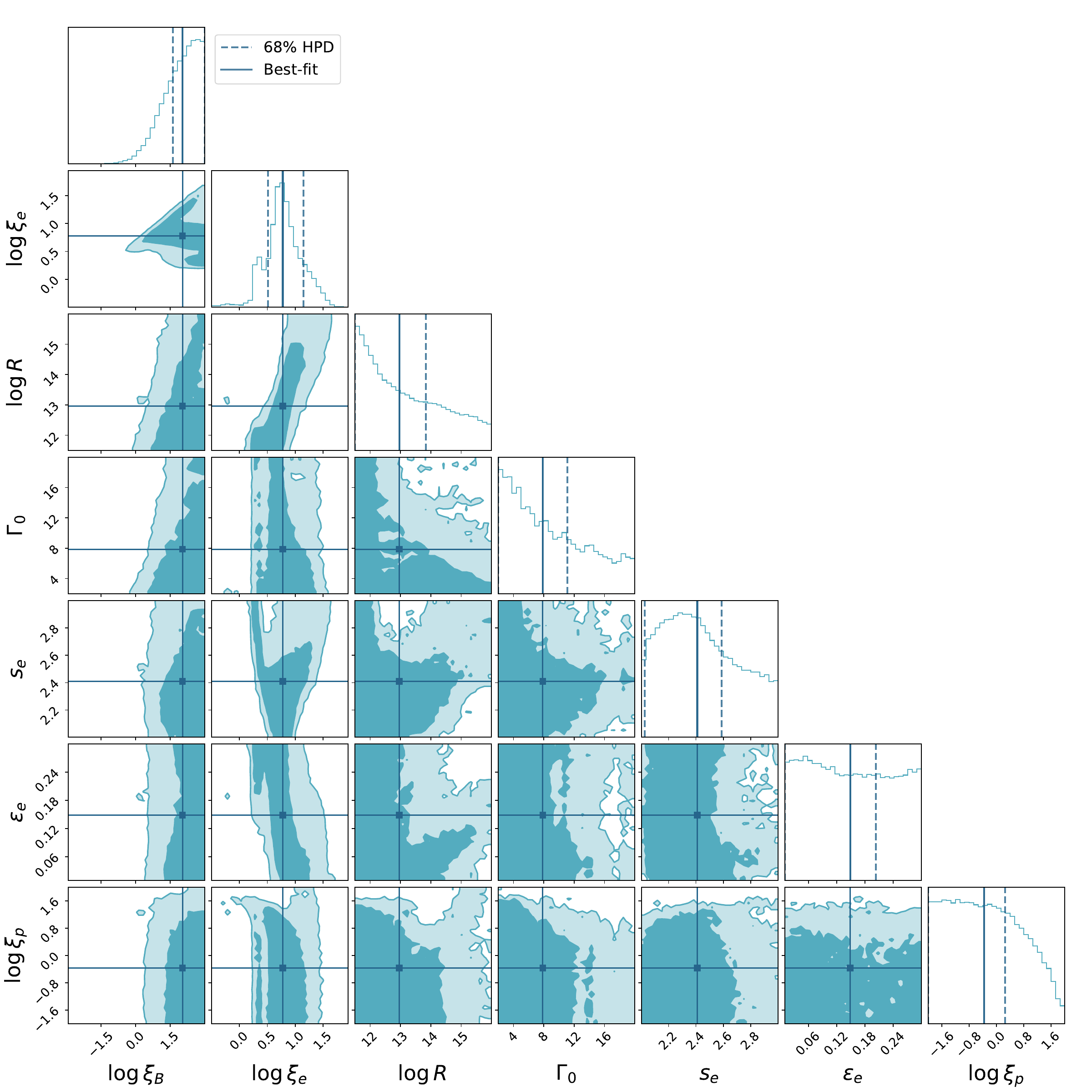} 
\includegraphics[width=0.4\linewidth]{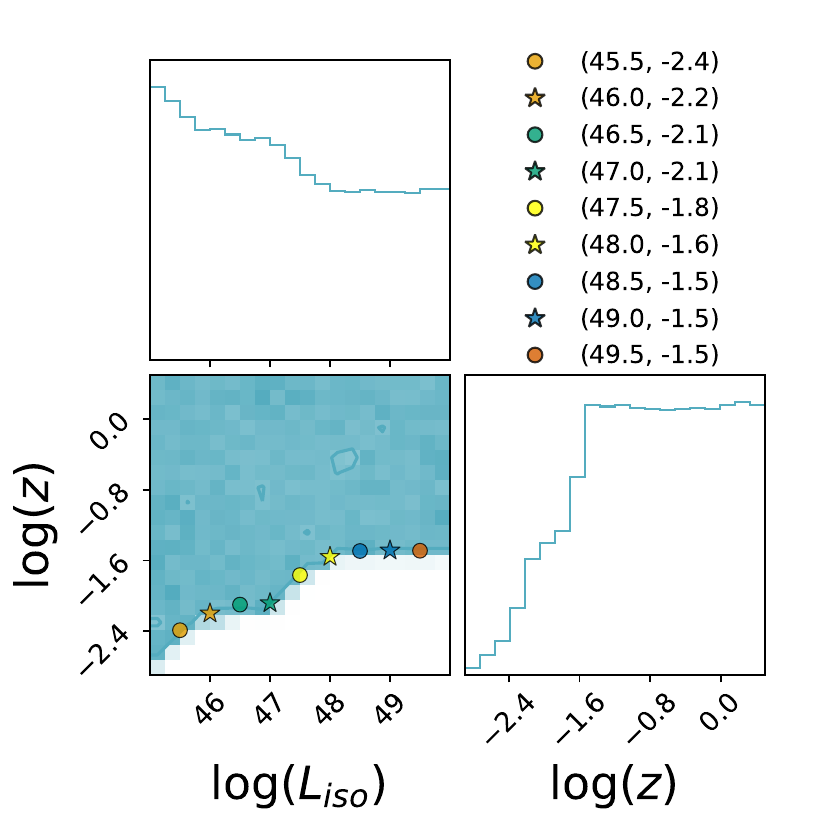}
\caption{Left: Confidence contours of the joint spectral fitting parameters for the prompt phase. Crossing points indicate the medians of the post-burn-in MCMC samples, while dashed lines mark the 1$\sigma$ HPD interval of the marginalized 1D posterior distribution; values and priors are provided in Table \ref{tab:prompt_bfp}. Right: Parameter space of isotropic gamma-ray luminosity ($L_{\gamma,\rm{iso}}$) and redshift ($z$) consistent with \fermilat{} upper limits under GRB~100316D configuration. Markers represent parameter points along the 95\% confidence level boundary used to illustrate the predicted flux in Fig.~\ref{fig:km3_fit}. Coordinates for these boundary samples are provided in the legend in the upper right panel.}
\label{fig:prompt_contour} \end{figure*}

\begin{figure}[hbt] \centering \includegraphics[width=0.48\linewidth]{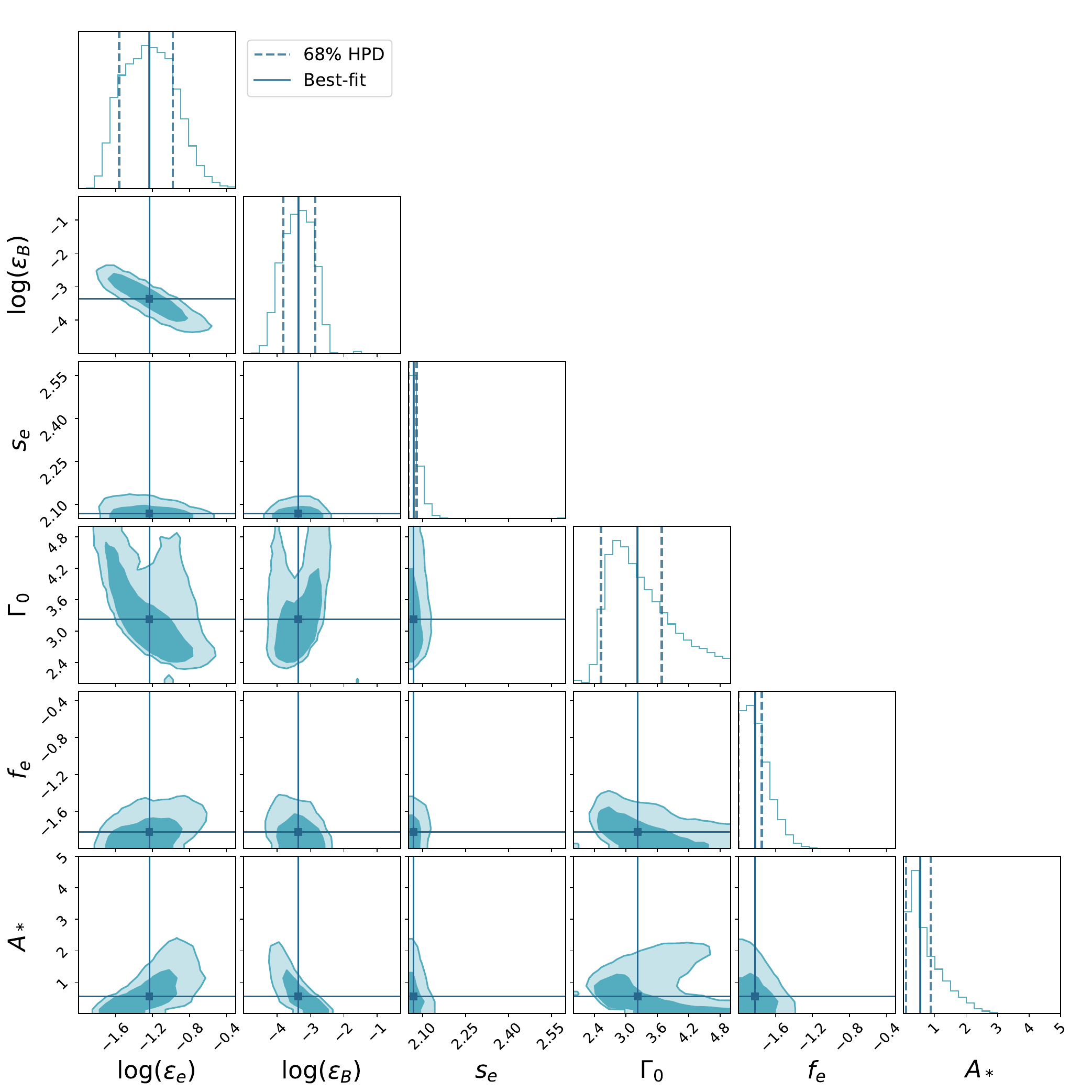} 
\includegraphics[width=0.48\linewidth]{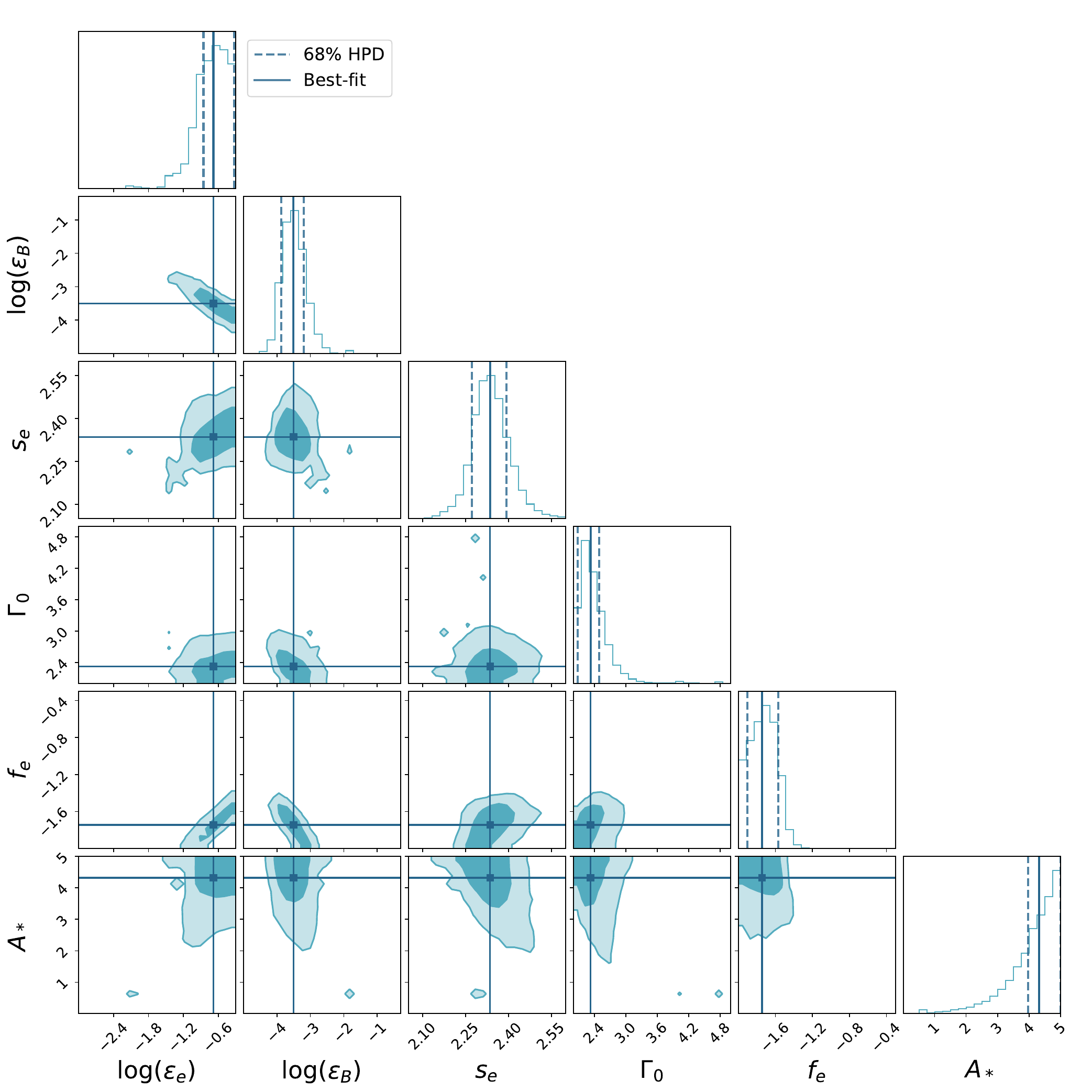} 
\caption{Confidence contours of the spectral fitting parameters for GRB~060218 (left) and GRB~100316D (right). Solid lines indicate the medians of the post-burn-in MCMC samples, while dashed lines mark the 1$\sigma$ HPD interval of the marginalized 1D posterior distribution; values and priors are provided in Table \ref{tab:afterglow_bfp}.} \label{fig:afterglow_contours} \end{figure}

The confidence contours for the prompt phase joint fit of SBO-like GRBs~060218 and 100316D LL~GRB are presented in Fig.~\ref{fig:prompt_contour} (left panel). The plots in Fig.~\ref{fig:afterglow_contours} show the posterior contours for fitting to the observed afterglow light curves of each event. In the right panel of Fig.~\ref{fig:prompt_contour}, the physically allowed parameter space of $L_{\gamma,\mathrm{iso}}$ and $z$ for KM3-230213A is constrained by the \fermilat{} upper limits, assuming the source has the properties of GRB~100316D.

The posterior distributions for several parameters in our analysis (see Fig.~\ref{fig:prompt_contour} and ~\ref{fig:afterglow_contours}) exhibit clustering near the adopted prior boundaries. These boundaries are not arbitrary search windows but are defined by physical requirements as detailed in Sec.~\ref{sec:model}. The clustering suggests that the data favor values at the bounds supported by standard theory and previous studies. 
It represents a regime where the current observational data do not provide information beyond the established theoretical floor. Breaking these parameter degeneracies will require higher precision multi-messenger data. Notably, mildly extending the prior ranges produces a negligible impact on the overall analysis outcomes.

\bibliography{main}
\bibliographystyle{aasjournalv7}
\end{document}